\documentclass[12pt]{article}
\usepackage{latexsym,amsmath,amssymb,}
\usepackage{graphicx}
\usepackage{ulem}
\usepackage{dsfont}

\begin{document}

\title{Solitons and Black Holes in a Generalized Skyrme Model with Dilaton-Quarkonium field}

\author{Daniela D. Doneva$^{1,2}$\thanks{E-mail: ddoneva@phys.uni-sofia.bg}
\\{\footnotesize  ${}^{1}$Deptartment of Astronomy,
                Faculty of Physics, St.Kliment Ohridski University of Sofia}\\
                {\footnotesize  5, James Bourchier Blvd., 1164 Sofia, Bulgaria }\\\\[-3.mm]
      {  \footnotesize ${}^{2}$ Theoretical Astrophysics, Eberhard-Karls University of T\"ubingen, T\"ubingen 72076, Germany }\\\\[-3.mm]
 Ivan Zh. Stefanov$^{3}$ \thanks{E-mail: izhivkov@tu-sofia.bg}
 \\ [2.mm]{\footnotesize{${}^{3}$ Department of Applied Physics, Technical University of Sofia,}}\\ [-1.mm]{\footnotesize{8, Kliment Ohridski Blvd.,
 1000
Sofia, Bulgaria}}\\\\[-3.mm]
Stoytcho S. Yazadjiev$^{4}$ \thanks{E-mail:
yazad@phys.uni-sofia.bg}\\
      {  \footnotesize ${}^{4}$Department of Theoretical Physics,
                Faculty of Physics, St.Kliment Ohridski University of Sofia}\\
{\footnotesize  5, James Bourchier Blvd., 1164 Sofia, Bulgaria }\\\\[-3.mm]
}

\date{}

\maketitle

\begin{abstract}
Skyrme theory is among the viable effective theories which emerge from low-energy limit of quantum chromodynamics. Many
of its generalizations include also a dilaton. Here we find new self-gravitating solutions, both solitons and black
holes, in a Generalized Skyrme Model (GSM) in which a dilaton is present.  The investigation of the properties of the
solutions is done numerically. We find that the introduction of the dilaton in the theory does not change the picture
qualitatively, only quantitatively. The model considered here has one free parameter more than the Einstein-Skyrme
model which comes from the potential of the dilaton. We have applied also the turning point method to establish that
one of the black-hole branches of solutions is unstable. The turning point method here is based on the first law of
black-hole thermodynamics a detailed derivation of which is given in the Appendix of the paper.
\end{abstract}

\noindent PACS numbers: 04.20.-q, 04.40.-b, 04.70.-s, 12.39.Dc \\



\sloppy
\section{Introduction}
Quantum chromodynamics (QCD) is a very successful theory in the regime of high energies where the relevant degrees of
freedom are quarks and gluons. This is the so-called asymptotic freedom phase. In the low-energy regime, however,  QCD
is extremely hard to treat since in that regime it is nonperturbative. As it was shown by t'Hooft \cite{tHooft1} QCD
can be treated perturbatively in the approximation of large color number $N_c\rightarrow \infty$. Within that
approximation QCD is equivalent to a local field theory of mesons and ``glueballs''. Latter, Witten \cite{Witten1,
Witten2} showed that  baryons can be represented by topological solutions -- solitons, in an effective theory of
mesons. Such an effective theory had actually been proposed much earlier by Skyrme \cite{Skyrme61, SkyrmeBook}. In his
theory baryons emerge as topological solitons, the so-called skyrmions. Their conserved topological charge, the winding
number, is identified as the baryon number.

Skyrme theory has proved successful in the description of the static properties of baryons but it has also many
deficiencies. Among them are the large masses and the missing intermediate-range nucleon-nucleon attraction which is
vital for the formation of nuclei. Different generalizations of the theory have been suggested to remedy these
deficiencies. One of the approaches is to add higher order derivative terms with the proper sign in the Lagrangian.
Another approach is the inclusion of additional fields in the theory. The scale invariance and the trace anomaly of the
underlying QCD theory in the low-energy limit leave place for the inclusion of scalar particles. For that reason many
of the suggested generalizations of Skyrme theory include a dilaton. The dilaton has considerable contribution to the
intermediate-range attractive forces \cite{Cottingham,Schwesinger, Riska, Kalbermann1}. The Lagrangian of a Generalized
Skyrme Model (GSM) including a dilaton has been derived from QCD in the low-energy regime in \cite{Adrianov86,
Adrianov87, Nikolaev92}.

Skyrme theory coupled to gravity to our knowledge has been considered for the first time by Luckock and Moss
\cite{Luckock}. They studied black-hole solutions. Self-gravitating skyrmions have been studied  by Glendenning,
Kodama, and Klinkhamer \cite{Glendenning88} in the approximation of large baryon numbers. They found that gravitating
skyrmions cannot be applied as models of baryon stars  since they are energetically unstable and disintegrate into
separate nonbounded solitons. Self-gravitating solitons with baryon number one have been investigated numerically by
Droz, Heuseler, and Straumann \cite{DrozHeusler91sol}. They have studied not only regular solutions but also black
holes. Their motivation was to find stable black-hole solutions which serve as counterexamples of the no-hair
conjecture. The black holes that they obtained have nontrivial Skyrme hair. The energy density of the matter field (the
Skyrme field) decays rapidly with the increase of the distance between the observer and the source and a distant
observer cannot distinguish them from the Schwarzschild black hole. In other words the Skyrme field does not contribute
to new independent, asymptotic charges. The authors found that these solutions are stable against spherically symmetric
perturbations \cite{HeuslerDroz91stab, HeuslerDrozStabLin}. Soon after that, Bizon and Chmaj \cite{Bizon92} found
another branch of solutions in the same model which turned out to be unstable. The stability of Einstein-Skyrme (ES)
solitons and black holes has been studied also in \cite{ Bizon07, Zajac09, Zajac10}. The Einstein-Skyrme black holes
have been generalized to the case of asymptotically anti-de Sitter spacetime  in \cite{Shiiki2005AdSa, Shiiki2005AdSb}.
Solitons and black holes in $SU(3)$ ES theory can be found in \cite{Kunz95}. The interior of the ES black holes has
been studied in \cite{TTM3}.

All of the gravitating solutions discussed so far are spherically symmetric and are based on the so-called hedgehog
ansatz. Monopole black-hole skyrmions based on a nontopological ansatz for the Skyrme field have been constructed in
\cite{MossShiikiWinstanley}. Axisymmetric soliton and black-hole solutions are given in \cite{SawadoShiiki2005}.  The
so-called harmonic map ansatz, which is a generalization of the rational map ansatz, has been applied in \cite
{Theodora2004} to obtain axially symmetric $SU(3)$ gravitating skyrmions and in \cite{Kunz06a} to obtain
self-gravitating $SU(2)$ skyrmions with discrete symmetries -- platonic skyrmions. The same authors obtained also
solutions for spinning skyrmions \cite{Kunz06b}. Employing the rational map ansatz Piette and Probert constructed
self-gravitating configurations which, unlike the solutions obtained with the hedgehog ansatz \cite{Glendenning88},
remain bounded even for very large baryon numbers \cite{Piette07}\footnote{Another possibility for the stabilization of
the multisoliton stars is the introduction of a chemical potential \cite{Loewe}.}.

Self-gravitating objects in  the generalizations of the Skyrme model which include a dilaton have not been so intensely
studied. K\"{a}lbermann obtained an effective equation of state (EOS) for nuclear matter starting from a generalized
Skyrme Lagrangian in which a dilaton and a $\omega$ meson are included \cite{Kalbermann97} \footnote{The $\omega$ meson
has been considered also in \cite{Nappi, Cottingham2}  for the stabilization of the skyrmion without higher derivative
terms. }. In this approach nuclear matter is described as skyrmion fluid which consists of free skyrmions with baryon
number one immersed in a mean field background. With this EOS massive skyrmion stars were constructed in
\cite{Ouyed99}--\cite{Ouyed08} \footnote{A wonderful brief review on Skyrme theory and its applications can be found in
\cite{Ouyed04}} and they were compared to neutron stars based on modern equations of state. Astrophysical applications
of the solutions and present plausible skyrmion star candidates are also given there. It appears that skyrmion stars
are quite promising since they can explain the high masses of some stars that have been observed and cannot be
explained by some of the modern EOS. Massive stars in theories with stiff EOS, such as Skyrme's, have been discussed
from an observational perspective in \cite{Popov}.

To our knowledge, self-gravitating soliton and black-hole solutions coupled to a dilaton have not been studied so far.
The purpose of the present paper is to make a step in that direction. We generalize the paper of Bizon and Chmaj
\cite{Bizon92} by adding a dilaton in the model. In particular we study self-gravitating solutions in the GSM proposed
in \cite{Nikolaev92}, \cite{Nikolaev00}.

The paper is organized as follows. The model is presented briefly in Sec. \ref{model}. The results of the numerical
investigation of the solutions are presented in Sec. \ref{Results} in which subsection \ref{Solitons} is dedicated to
the solitons, and subsection \ref{Black_Holes} -- to the black-hole solutions. In Sec.
\ref{Stability_and_thermodynamics} the stability of the  GSM black-hole solutions is studied with the application of
Poincare's turning point method. Here the turning point method is based on the first law of black-hole thermodynamics a
detailed derivation of which is given in the Appendix. The results are summarized  in the Conclusion.

\section{Basic equations}\label{model}
We start with the following action:
\begin{equation}
S=\int{d^4x \sqrt{-g} \left(-\frac{R}{16\pi G} + L_M \right)}. \label{eq:action}
\end{equation}
For the matter sector $L_M$ we consider the effective GSM Lagrangian  \cite{Nikolaev92} in which the derivatives have
been substituted with covariant derivatives
\begin{eqnarray}
&&L_M = \frac{1}{4} f_\pi^2~ \exp(-2\sigma)~ {\rm Tr}[\nabla_\mu U \nabla^\mu U^+]  + \frac{N_f f_\pi^2}{4}~ \exp(-2\sigma)~ g^{\mu\nu} \partial_\mu \sigma
\partial_\nu \sigma
\label{eq:Lagrang}\\ \notag \\
&& \hspace{0.5cm} + \frac{1}{32 e^2} {\rm Tr}[(\nabla_\mu U) U^{+},(\nabla_\nu U)
U^{+}]^2  + \tilde{V}(\sigma) . \notag
\end{eqnarray}
Here $U$ is the SU(2) chiral field, $\sigma$ is the dilaton, $\nabla_\mu$ is the covariant derivative with respect to the metric
$g_{\mu\nu}$, $f_\pi$ is the pion decay
constant, $e$ is the Skyrme constant, $C_g$ is the gluon condensate, $N_f$ is the number of flavors and $\varepsilon =
8N_f/(33-2N_f)$.

The first two terms in (\ref{eq:Lagrang}) are the kinetic terms for the chiral and the dilaton fields. The third term is the
one introduced by Skyrme for the stabilization of the soliton solutions and the last one is the potential of the dilaton field given by
\begin{equation}
\tilde{V}(\sigma) = -\frac{C_g N_f}{48} \left[ \exp(-4\sigma) - 1 + \frac{4}{\varepsilon} (1-\exp(-\varepsilon
\sigma))
\right]. \label{eq:PotentOrig}
\end{equation}
As mentioned above the dilaton is introduced in the theory to restore scale invariance. It couples to those terms of
the Lagrangian density that break the scale invariance \cite{Riska,Schecter}. In the classical Skyrme theory the first
term breaks scale invariance while the stabilizing term is scale invariant.

We will restrict our study to static, spherically symmetric, asymptotically flat solutions. Hence, the following ansatz
for the metric can be used:
\begin{equation} ds^2 = A^2(r)\left(1-\frac{2 m(r)}{r}\right) dt^2 - \left(1-\frac{2
m(r)}{r}\right)^{-1} dr^2 - r^2(d\theta^2 + \sin^2{\theta} d\varphi^2) \label{eq:metric}
\end{equation}
and the hedgehog ansatz for the chiral field
\begin{equation}
U=\exp[\mathbf{\tau}\cdot \hat{\mathbf{r}} F(r)],
\end{equation}
where $\tau$ are the Pauli matrices and $\hat{\mathbf{r}}$ is a unit radial vector. It is also useful to introduce a new
function $\Phi$ for the scalar field instead of $\sigma$, defined by
\begin{equation}
\Phi=\exp(-\sigma).
\end{equation}

The reduced field equations following from the action (\ref{eq:action}) are
\begin{eqnarray}
&&\hspace{4cm}m'= -\frac{r^2}{4} {\cal L}_m, \label{eq_m}\\ \notag \\
&& \hspace{3cm}A' = {1\over 2} \left({u\over r} F\,'^{\,2}+ a\,r\,\tilde{N} \Phi\,'^{\,2} \right)A\label{eqA},\\ \notag \\
&& \hspace{1cm} (f\,A\,u\,F\,' )\,'-{1\over 2}\,f\,A\,F\,'^{\,2}\,{\partial u\over \partial F}-{1\over 2}\,A\,{\partial v \over \partial
F}=0,
\label{eqF}\\ \notag \\
&& a\,\tilde{N}\,(r^2\,f\,A\,\Phi\,' )\,'-{1\over 2}\,f\,A\,F\,'^{\,2}\,{\partial u\over \partial \Phi}-{1\over 2}\,A\,{\partial v \over \partial
\Phi}+{1\over 2}\,r^{\,2}\,A\,{\partial V\over \partial \Phi}=0,\label{eqPhi}
\end{eqnarray}
where
\begin{eqnarray}
&&\hspace{2cm}{\cal L}_m=-f {u\over r^2} F\,'^{\,2} - {v\over r^2}- a \tilde{N}f \Phi\,'^{\,2}+V(\Phi),\\\notag \\
&& \hspace{1.2cm}V(\Phi)=16 \pi G \tilde{V}(\Phi)=-\gamma \tilde{N}\left[\Phi^4-1+{4\over
\varepsilon}\left(1-\Phi^{\varepsilon}\right)\right],
\label{potential} \\ \notag \\
&&f=1-\frac{2 m(r)}{r}, \hspace{0.2cm}u=a\,r^2\Phi^2+2\,b\,\sin^2{F}, \hspace{0.2cm}v=\left(2\,a\,\Phi^2+b\,{\sin^2{F}\over
r^2}\right)\sin^2{F}.
\end{eqnarray}
We have introduced the following constants:
\begin{equation}
a = 8\pi G f^2_{\pi}, ~~ b=8\pi G \frac{1}{e^2}, ~~ \gamma=2 \pi G \frac{C_g}{3}, ~~\tilde{N}={N_f\over2},
\end{equation}
where $a$ and $\tilde{N}$ are dimensionless.
It can be easily seen that the system of field equations (\ref{eq_m})--(\ref{eqPhi}) does not admit in general solutions
with trivial dilaton field $\sigma=0$ ($\Phi=1$).

For the numerical treatment of the system it is convenient to work with the following dimensionless variables and
parameters:
\begin{eqnarray}
&&x = e f_\pi r, ~~~\mu(x) = e f_\pi m, ~~~ D_{\rm eff} =  {\gamma \tilde{N}\over 2~a~ e^2 f_\pi^2 } .
\end{eqnarray}
The dimensionless field equations can be written in a form which depends only on the two coupling parameters $a$ and
$D_{\rm eff}$~\footnote{The parameter $a$ is two times bigger than the parameter $\alpha$ used in \cite{Bizon92} and
\cite{DrozHeusler91sol} and the parameter $D_{\rm eff}$ is chosen to be the same as in
\cite{Nikolaev92},\cite{Nikolaev00}.} and on the parameter connected to the number of the flavors $\tilde{N}$. For the
number of flavors we fixed the value  $N_f=2$, so $\tilde{N}=1$.

We will consider two classes of solution to the above field equations -- solitons and black holes that can be obtained
by imposing the appropriate  boundary conditions. In the next section we will describe in detail the required boundary
conditions and the obtained results.

\section{Numerical results}\label{Results}

\subsection{Solitons}\label{Solitons}
The domain of integration for the soliton solutions is $x \in [0,\infty)$ and the requirement that the functions should
be regular at the origin leads to the following expansions of the functions at $x\rightarrow 0$:

\begin{eqnarray}
&&\left.\mu(x)\right|_{x\rightarrow 0} = \Big[\frac{1}{6} a D_{\rm eff} \left(\Phi_0^{~4} - 1 +
\frac{4}{\varepsilon}(1-\Phi_0^{~\varepsilon})\right) +  \label{eq:BC_soliton_m(0)} \\ \notag \\ &&\hskip 2.2cm\frac{1}{4} a {F'_0}^{2} \left(
\Phi_0^{~2} +  {F'_0}^{2}\right) \Big]
x^3   + {\rm O}(x^5), \notag \\ \notag \\
&&\left.A(x)\right|_{x\rightarrow 0} = A_0 + \frac{1}{4} a A_0 {F'_0}^2 \left( \Phi_0^{~2} +  2{F'_0}^{2}\right) x^2 + {\rm
O}(x^4),
\label{eq:BC_soliton_A(0)} \\
\notag \\
&& \left.F(x)\right|_{x\rightarrow 0} =n \pi + {F'_0}x + {\rm O}(x^3), \label{eq:BC_soliton_F(0)} \\ \notag \\
&&\left.\Phi(x)\right|_{x\rightarrow 0} =\Phi_0 +
\left[\frac{2D_{\rm eff}}{3\tilde{N}}(\Phi_0^{~3}-\Phi_0^{~\varepsilon-1}) +\frac{1}{2\tilde{N}}\Phi_0 {F'_0}^{2} \right]x^2 + {\rm O}(x^4),  \label{eq:BC_soliton_Phi(0)}
\end{eqnarray}
where $n$ is an integer number, $A_0$ and $\Phi_0$ are the values of the corresponding functions at $x=0$ and $F'_0$
is the first derivative of $F$ with respect to $x$ at $x=0$. The shooting parameters are $A_0$, $\Phi_0$ and $F'_0$.

The asymptotic flatness of the solutions requires \footnote{The boundary conditions at infinity are also manifested by the asymptotic expansion of
the
functions at infinity (\ref{eq:Assympt_MA_inf}) and (\ref{eq:Assympt_FPhi_inf}).}
\begin{eqnarray}
&& f(x)|_{x\rightarrow \infty} = 1,\hspace{0.6cm} A(x)|_{x\rightarrow \infty} = 1,\hspace{0.6cm} \Phi(x)|_{x\rightarrow \infty}=1,\hspace{0.6cm} F(x)|_{x\rightarrow \infty}
=0. \label{eq:BC_inf}
\end{eqnarray}
The last boundary condition for the function $F$ corresponds to the following requirement for the Skyrme chiral field:
$U(x)|_{x\rightarrow \infty}=1$. This effectively compactifies ${\rm R}^3$ to ${\rm S}^3$ and the chiral field can be
considered as a map from ${\rm S^3}$ to ${\rm SU}(2)$, i.e., a map from ${\rm S}^3$ to ${\rm S}^3$ since ${\rm SU}(2)$
is topologically ${\rm S}^3$. The third homotopy class of ${\rm SU}(2)$ is ${\rm Z}$ which means that every field
configuration is characterized by an integer winding number $B$. This winding number is usually interpreted as the
baryon number and is given by
\begin{equation}
B=\int{{\rm d}^3 x \sqrt{-g} B_0},
\end{equation}
where the topological current $B_\mu$ is
\begin{equation}
B_\mu = \frac{\epsilon_{\mu\nu\alpha\beta}}{24 \pi^2} {\rm Tr} [(U^{+}\nabla^\nu U)(U^{+}\nabla^\alpha U)(U^{+}\nabla^\beta
U)].
\end{equation}
For the hedgehog ansatz and for the imposed boundary conditions $B=n$.

\begin{figure}
    \begin{center}
    \includegraphics[width=7.5cm]{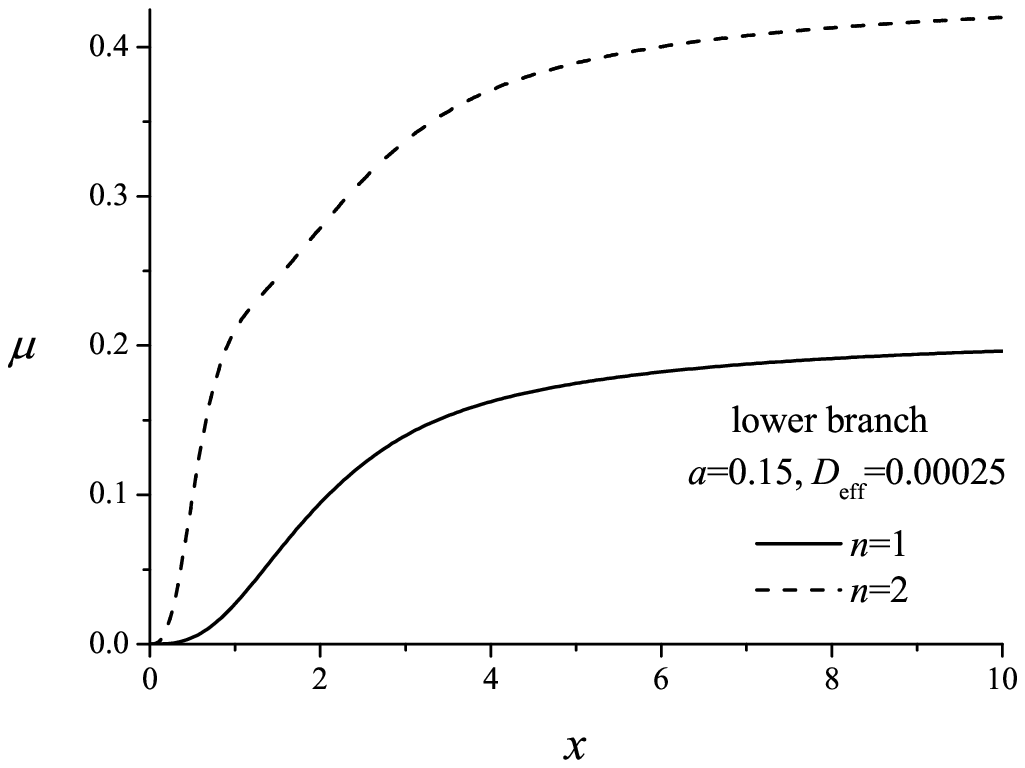}
    \includegraphics[width=7.5cm]{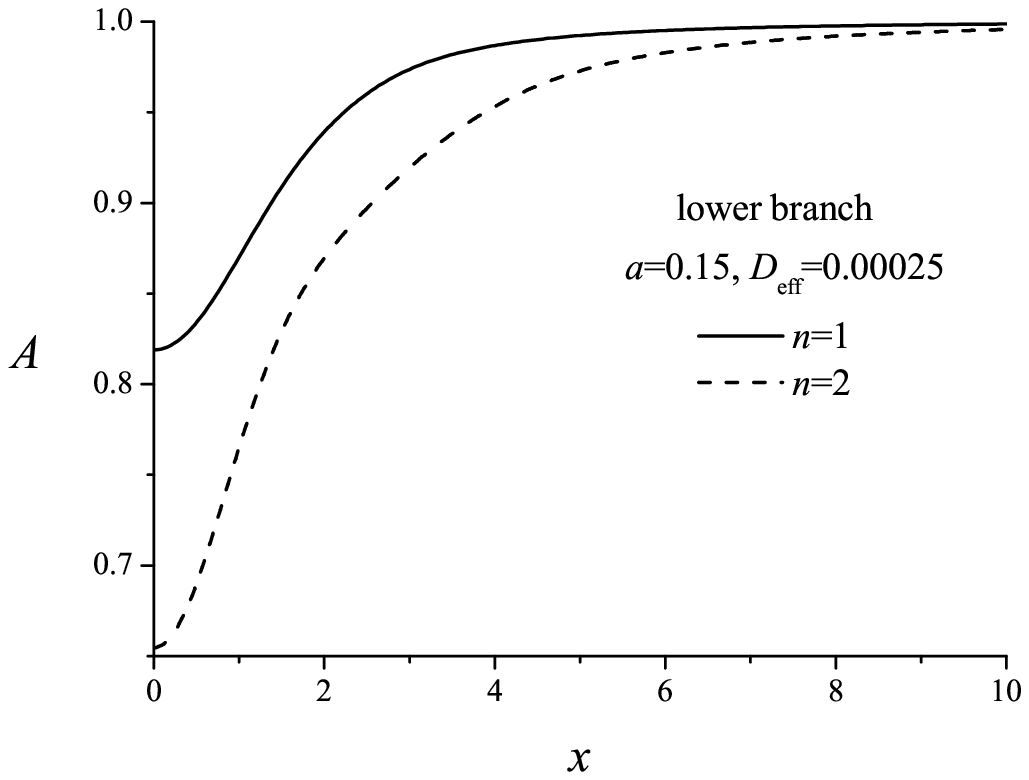}
    \caption{
    The metric functions $\mu(x)$ (left panel) and $A(x)$ (right panel) for soliton solutions with different value of the parameter $n$ where $D_{\rm eff}=0.00025$,
    $a=0.15$ and $\tilde{N}=1$. Solitons which belong only to the \textit{lower branch} of solutions are presented and the \textit{upper branch} of solutions have similar
    qualitative behavior (\textit{lower} and \textit{upper branch} are defined below).
        }
    \label{fig:mA(x)}
\end{center}
\end{figure}

\begin{figure}
    \begin{center}
    \includegraphics[width=7.5cm]{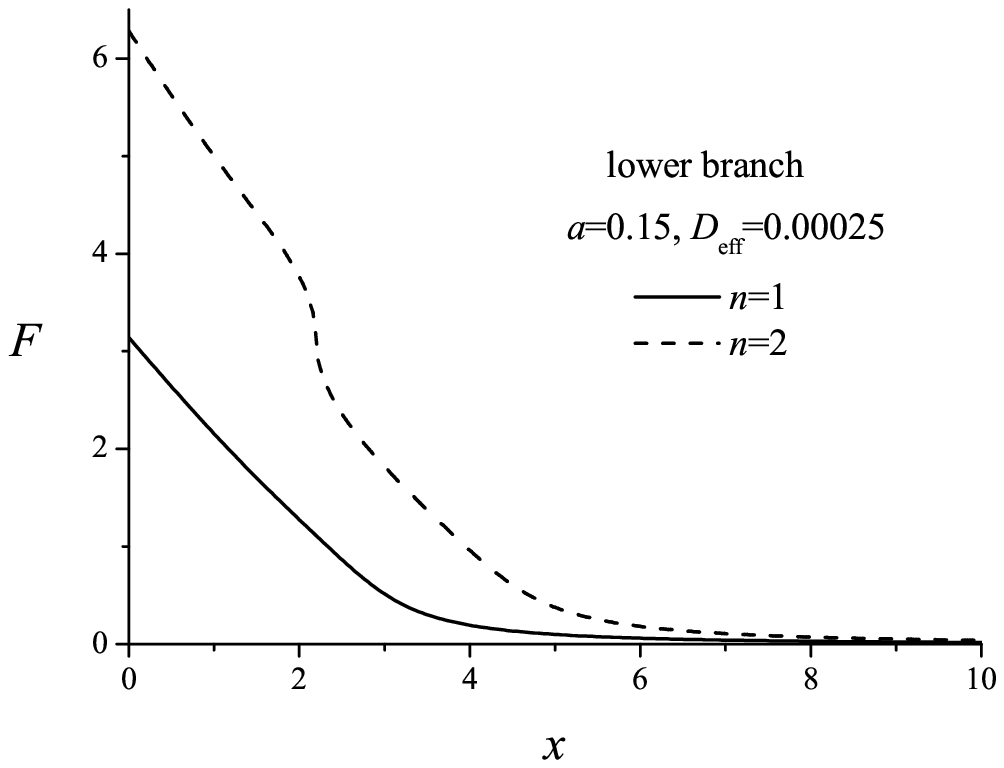}
    \includegraphics[width=7.5cm]{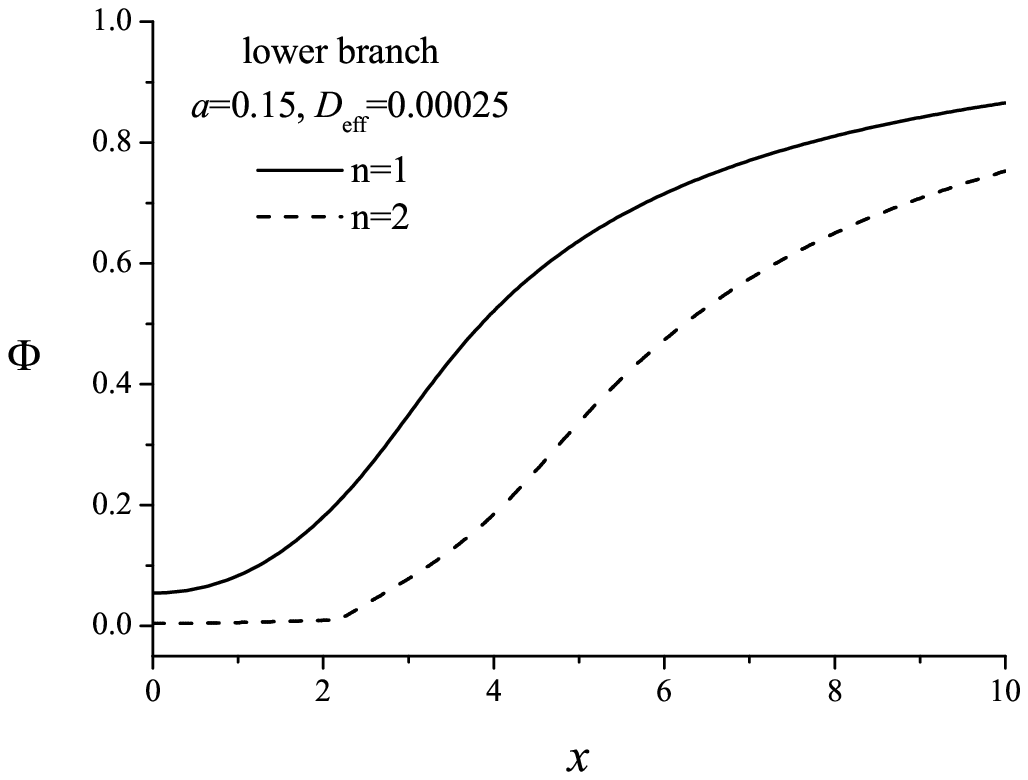}
    \caption{
    The skyrmion field $F(x)$ (left panel) and the dilaton $\Phi(x)$ (right panel) for  the same solutions as on Figure
    \ref{fig:mA(x)} are shown.
        }
    \label{fig:FPhi(x)}
\end{center}
\end{figure}

The field equations (\ref{eq_m})--(\ref{eqPhi}) together with the above boundary conditions are solved numerically
using the shooting method. On Figs.  \ref{fig:mA(x)} and \ref{fig:FPhi(x)} the metric functions $\mu(x)$ and $A(x)$,
the Skrymion field $F(x)$ and the scalar field $\Phi(x)$ are presented for several soliton solutions obtained for
$D_{\rm eff}=0.00025$, $a=0.15$, $\tilde{N}=1$ and for different values of the parameter $n$. As it can be seen the
profiles of the functions differ significantly for different values of $n$ and a general observation is that the number
of the ``steps'' of the function $F(x)$ is equal to $n-1$ \footnote{Such ``steps'' of the $F(x)$ function exist also in
the case when no scalar field is present.}.

\begin{figure}
    \begin{center}
    \includegraphics[width=7.5cm]{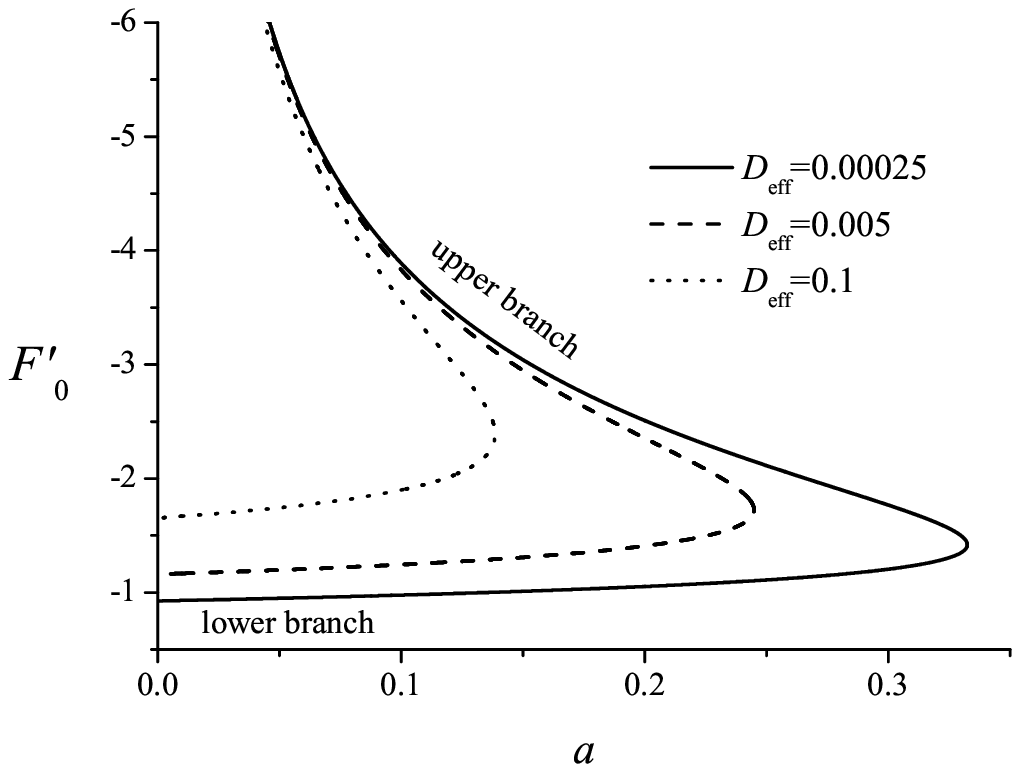}
    \includegraphics[width=7.5cm]{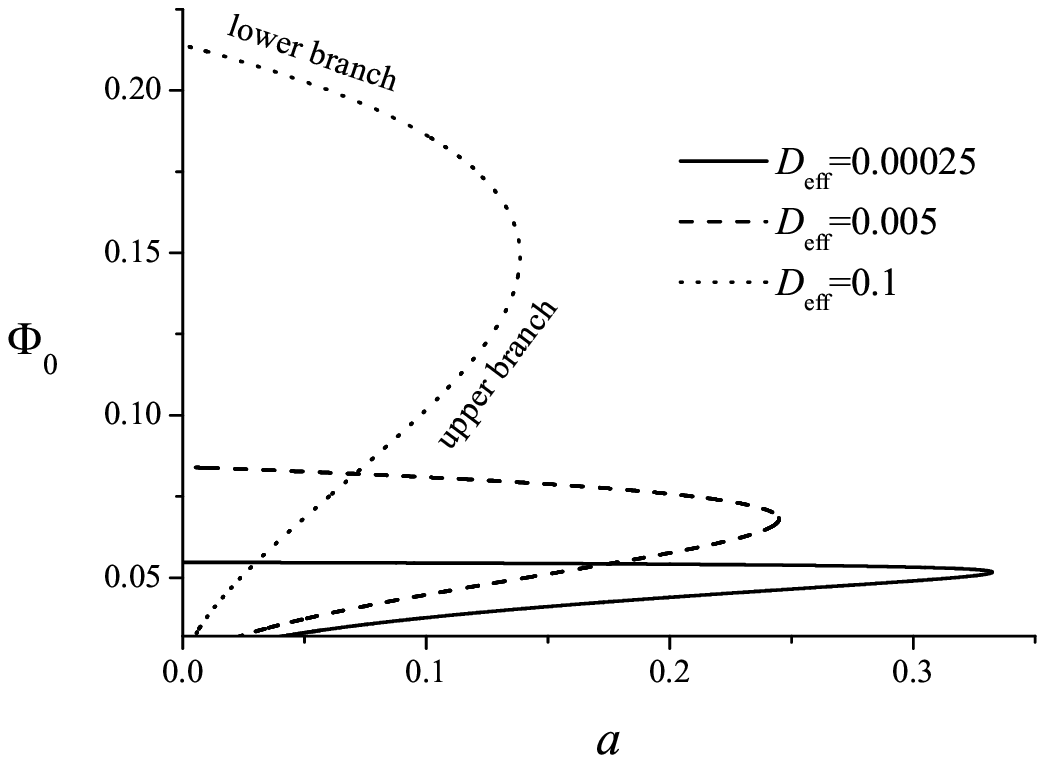}
    \caption{
    The shooting parameters $F'_0$ and $\Phi_0$ as functions of the coupling parameter $a$ for different $D_{\rm eff}$ are shown.
        }
    \label{fig:FprPhi(a)_var_eta}
\end{center}
\end{figure}

\begin{figure}
    \begin{center}
    \includegraphics[width=7.5cm]{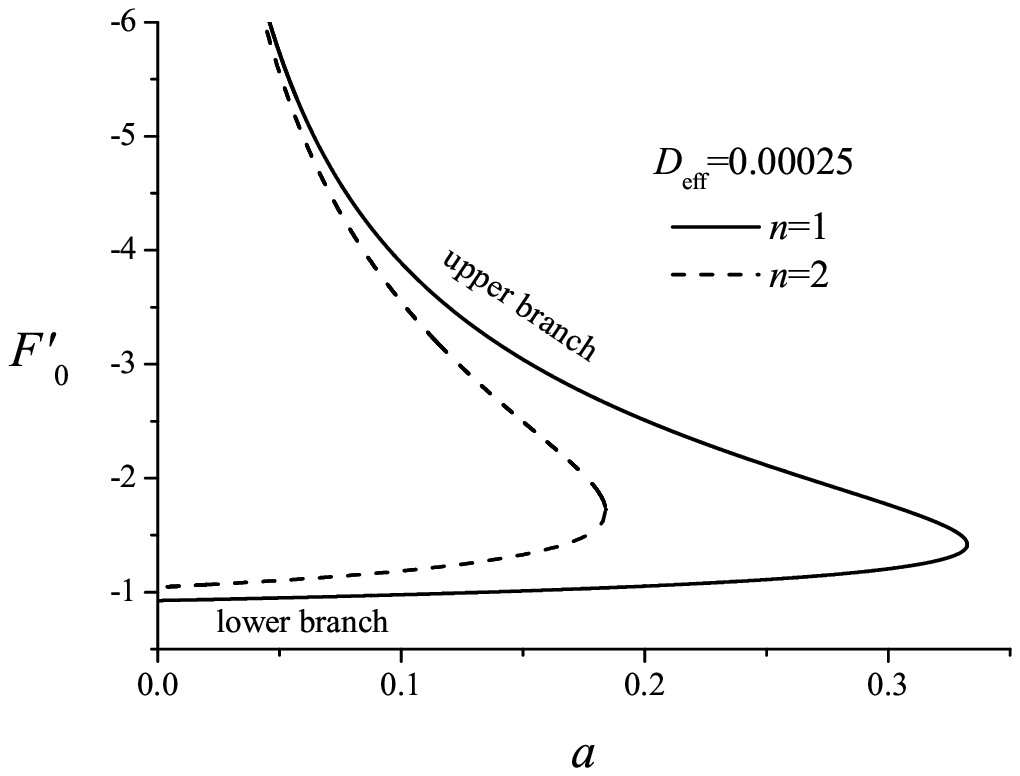}
    \includegraphics[width=7.5cm]{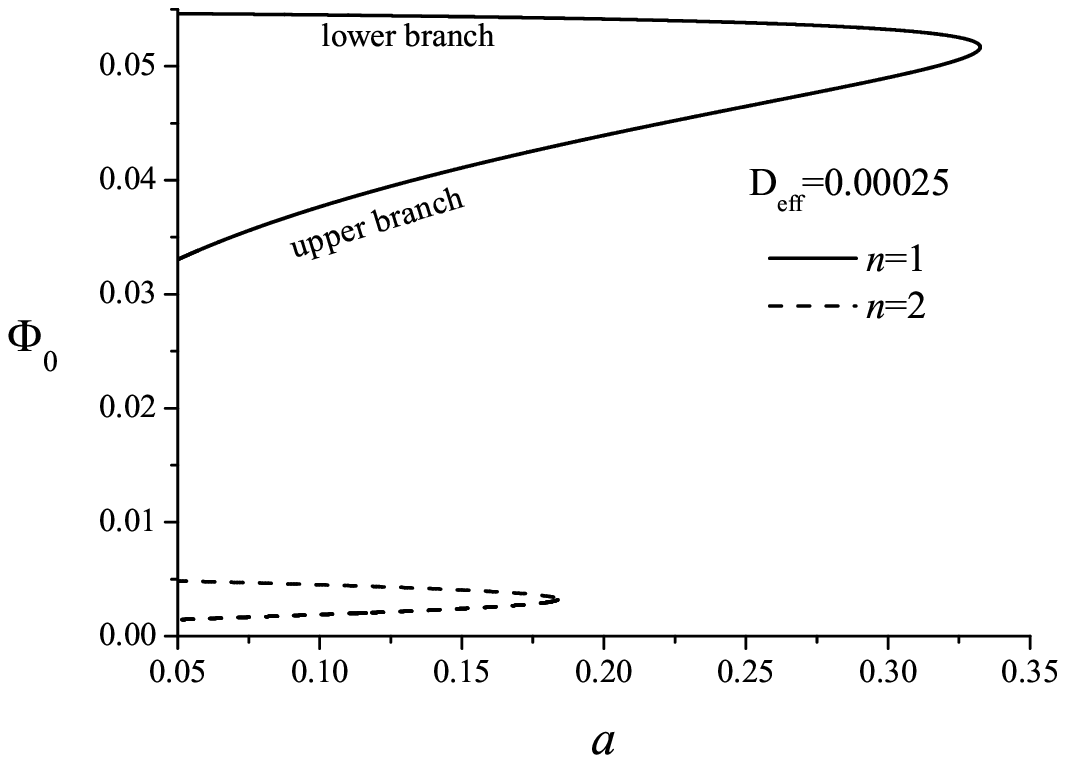}
    \caption{
    The shooting parameters $F'_0$ and $\Phi_0$ as functions of the coupling parameter $a$  for soliton with
    $n=1$ and $n=2$, where $D_{\rm eff}=0.00025$ and $\tilde{N}=1$.}
    \label{fig:Fpr(a)_var_n}
\end{center}
\end{figure}

Similarly to the case when the dilaton is not present, nonuniqueness of the solutions exits. For fixed values of the
input parameters two soliton solutions with different values of the shooting parameters exist which belong to the
so-called \textit{upper} and \textit{lower} branches of solutions\footnote{When we refer to \textit{upper} and
\textit{lower branch} of the soliton solutions, we use the convention from \cite{Bizon92} which is based on the $F'_0$
phase diagram. This is exactly opposite to the $\Phi_0$ phase diagram where the \textit{upper branch} is below the
\textit{lower branch}.}. These branches are presented on Fig. \ref{fig:FprPhi(a)_var_eta} where the shooting parameters
$F'_0$ and $\Phi_0$ are given as a function of the coupling parameter $a$  for several values of $D_{\rm eff}$. The
$F'_0(a)$ and $\Phi_0(a)$ dependences for different values of $n$ are given on Fig. \ref{fig:Fpr(a)_var_n} for $D_{\rm
eff}=0.00025$.

The obtained sequences of solutions are qualitatively very similar to the case without dilaton field. The main
difference is that the maximal value of the coupling parameter $a_{\rm max}$ for which soliton solutions exist depends
strongly on the coupling parameter $D_{\rm eff}$ and can be several times larger than in the case without scalar field.
An interesting observation is that when $D_{\rm eff}$ approaches the values considered in
\cite{Nikolaev92},\cite{Nikolaev00} (on the plots $D_{\rm eff}=0.1$), $a_{\rm max}$ approaches the values in the
Einstein-Skyrme model (ESM).

\subsection{Black Holes}\label{Black_Holes}
In the case of black holes the domain of integration is $x \in [x_H,\infty)$, where $x_H$ is the dimensionless radius of
the black-hole horizon. The presence of a horizon makes the black hole topologically trivial
because the three-dimensional space ${\rm R}^3$ is equivalent to ${\rm S^3}$ minus a ball and the map $U(x)$ is
topologically trivial. That is why the function $F$  does not have to be a multiple of $\pi$ at the black-hole horizon.

The boundary conditions at infinity are again deduced from the asymptotic flatness requirement
and they are the same as in the soliton case (\ref{eq:BC_inf}).
The boundary condition for the dimensionless metric function $\mu$ at the horizon is the standard one
\begin{eqnarray}
&&\mu(x_H) = \frac{x_H}{2}.
\end{eqnarray}
The regularization conditions at the horizon are
\begin{eqnarray}
&& \left.\frac{dF}{dx}\right|_{x\rightarrow x_H} = \frac{\sin{2F}}{x} \frac{\left(\Phi^2 +  \frac{\sin^2{F}}{x^2} \right)}{\left(\Phi^2 +
2\frac{\sin^2{F}}{x^2}
\right)} \times \\&& \notag \\
&& \hspace{2cm} \times \frac{1}{1-\frac{a}{2}\sin^2{F}\left(2\Phi^2 + \frac{\sin^2{F}}{x^2}\right) - x^2 a D_{\rm eff} \left[\Phi^4 - 1
+\frac{4}{\varepsilon}(1-\Phi^\varepsilon)\right]}, \notag
\end{eqnarray}
\begin{eqnarray}
&& \left.\frac{d\Phi}{dx}\right|_{x\rightarrow x_H} = \left(4x\frac{D_{\rm eff}}{\tilde{N}}(\Phi^3-\Phi^{\varepsilon-1}) + \frac{2}{\tilde{N}}\frac{\sin^2{F}}{x}
\Phi \right)\times \\&& \notag \\
&& \hspace{2cm} \times \frac{1}{1-\frac{a}{2}\sin^2{F}\left(2\Phi^2 + \frac{\sin^2{F}}{x^2}\right) - x^2 a D_{\rm eff} \left[\Phi^4 - 1
+\frac{4}{\varepsilon}(1-\Phi^\varepsilon)\right]}. \notag
\end{eqnarray}
Using the above boundary conditions we can conclude that in the case of the black holes we have again three shooting
parameters, $A_H$, $F_H$, and $\Phi_H$ which denote the values of the corresponding functions on the horizon.

\begin{figure}
    \begin{center}
    \includegraphics[width=7cm]{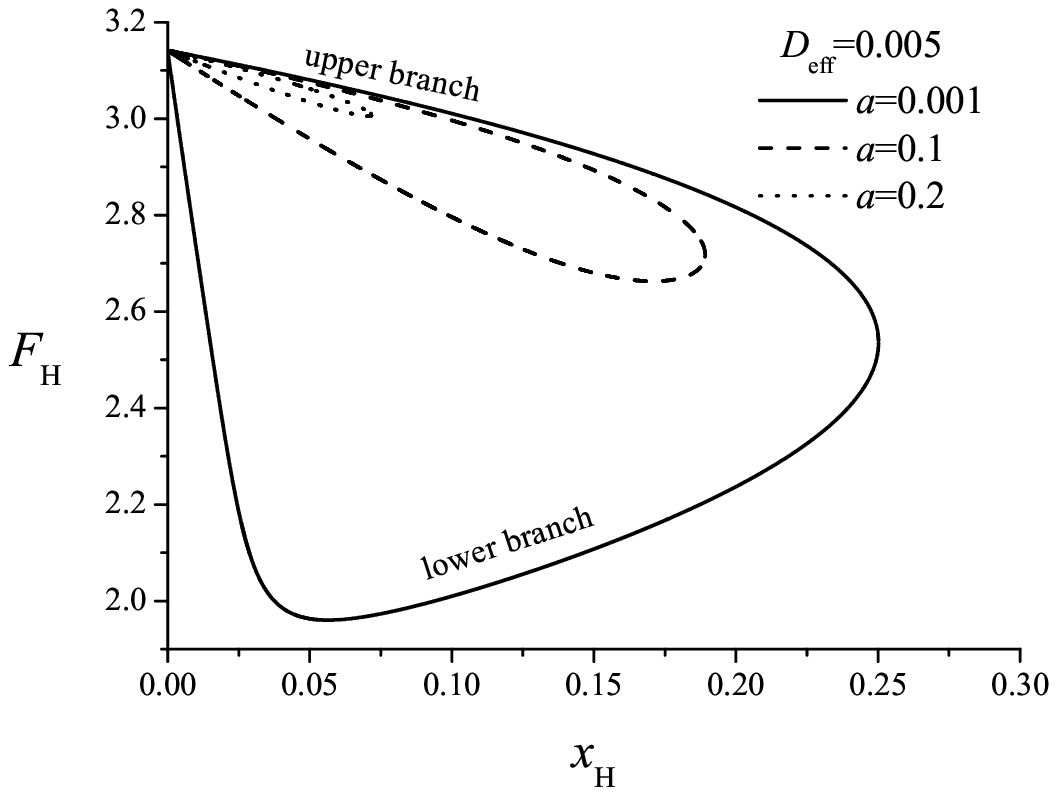}
    \includegraphics[width=7cm]{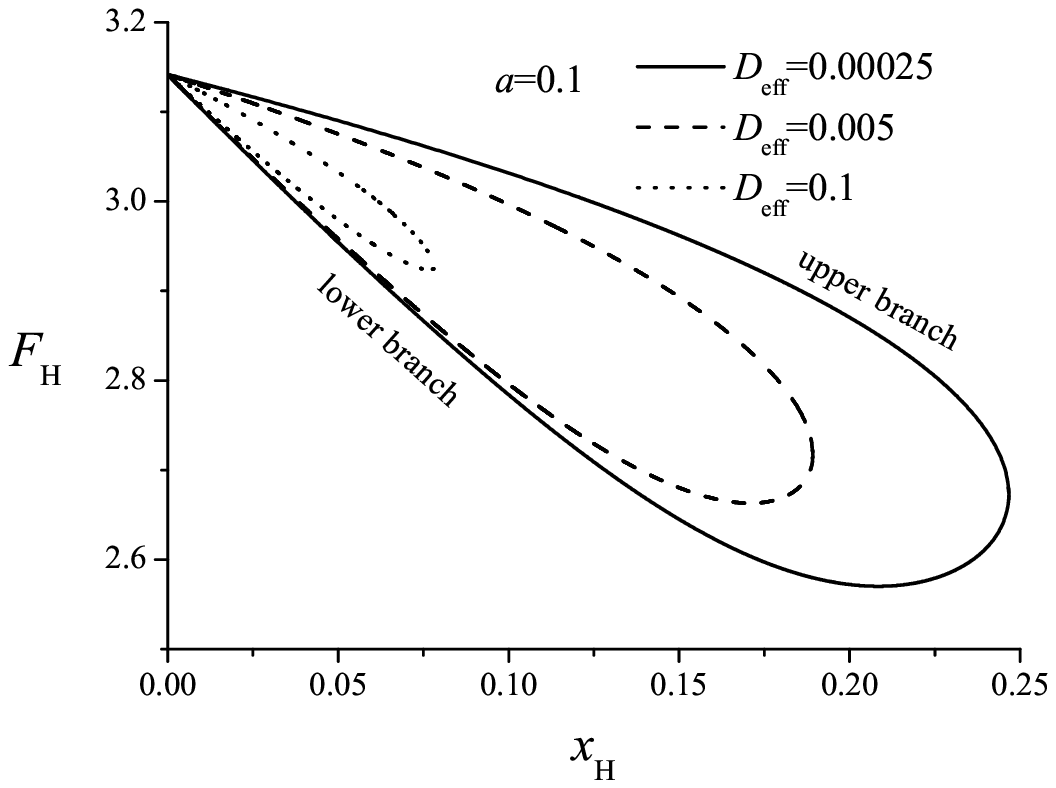}
    \caption{
    The shooting parameter $F_H$ as a function of the radius of the black-hole horizon $x_H$
    for several values of the coupling parameters. The results for $D_{\rm eff}=0.005$ and for various values of $a$ are
    shown on the left panel and the results for $a=0.1$ and for various valued of the parameter $D_{\rm eff}$ are shown on the
    right panel.
        }
    \label{fig:Frh(a)}
\end{center}
\end{figure}

\begin{figure}
    \begin{center}
    \includegraphics[width=7cm]{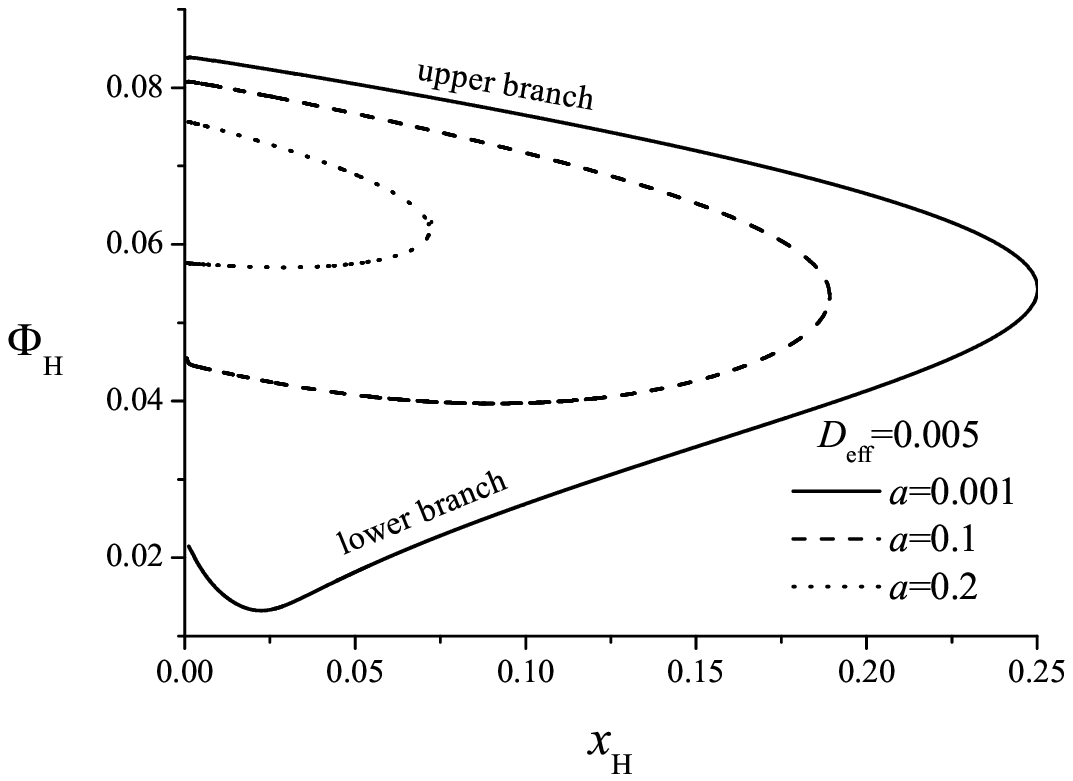}
    \includegraphics[width=7cm]{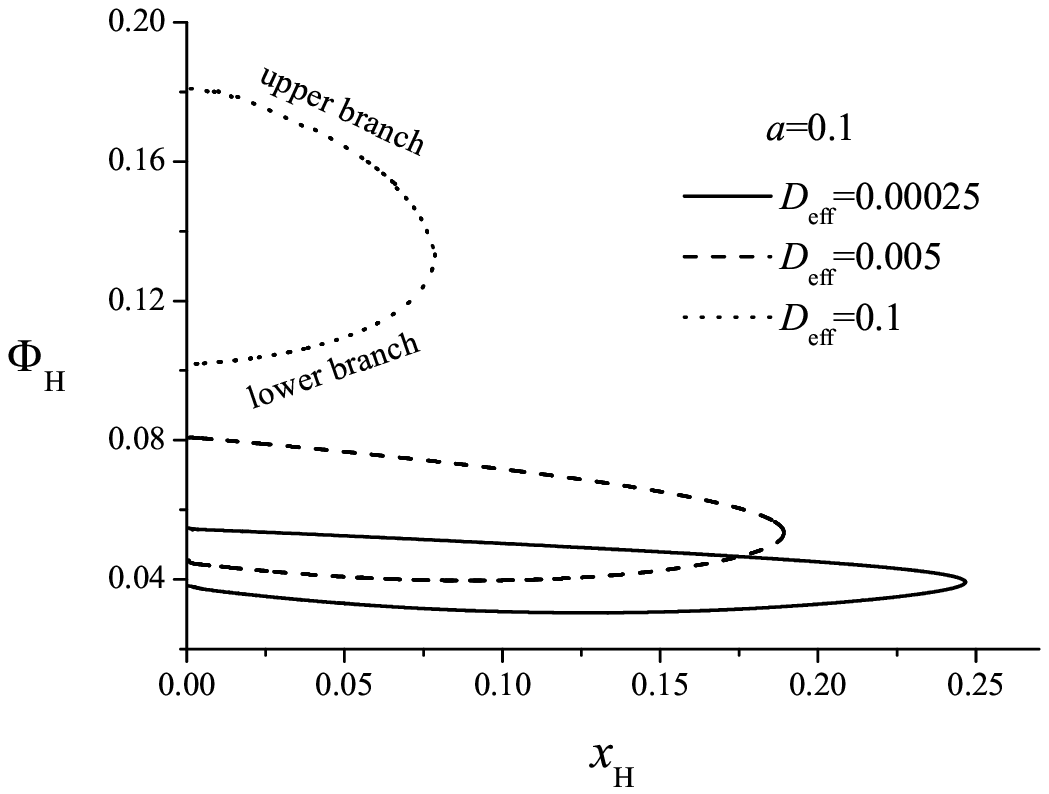}
    \caption{
    The shooting parameter $\Phi_H$ as a function of the radius of the black-hole horizon $x_H$
    for the same black-hole solutions as on Fig. \ref{fig:Frh(a)} is shown.
        }
    \label{fig:Phirh(a)}
\end{center}
\end{figure}

\begin{figure}
    \begin{center}
    \includegraphics[width=7cm]{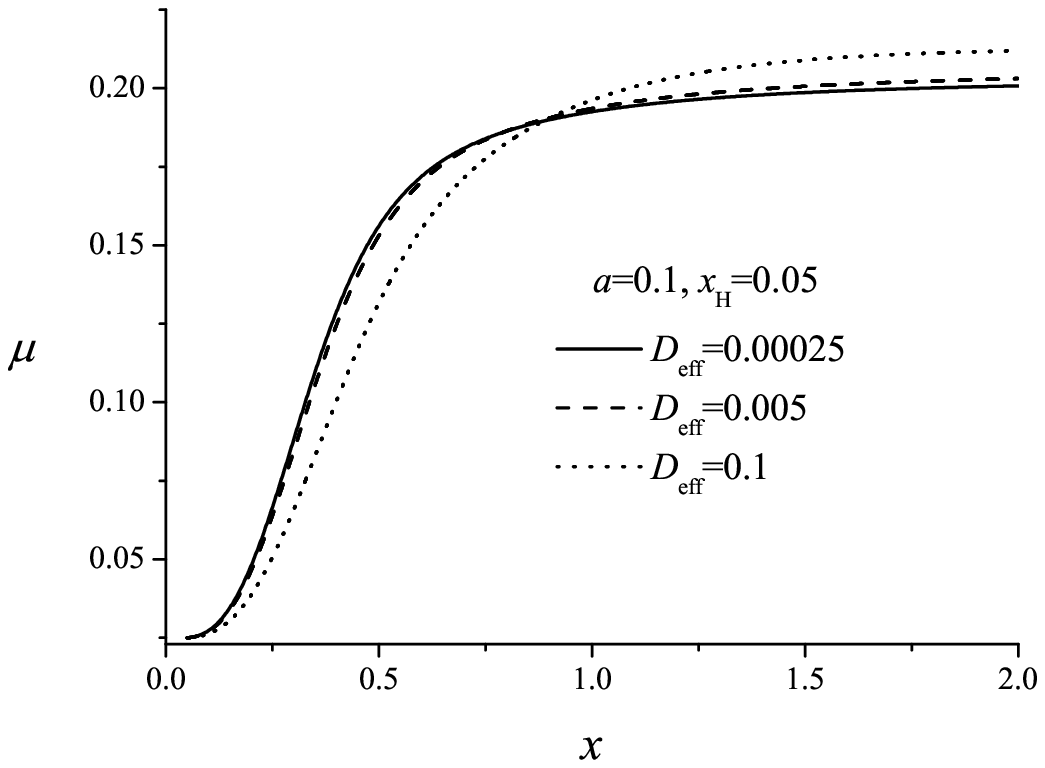}
    \includegraphics[width=7cm]{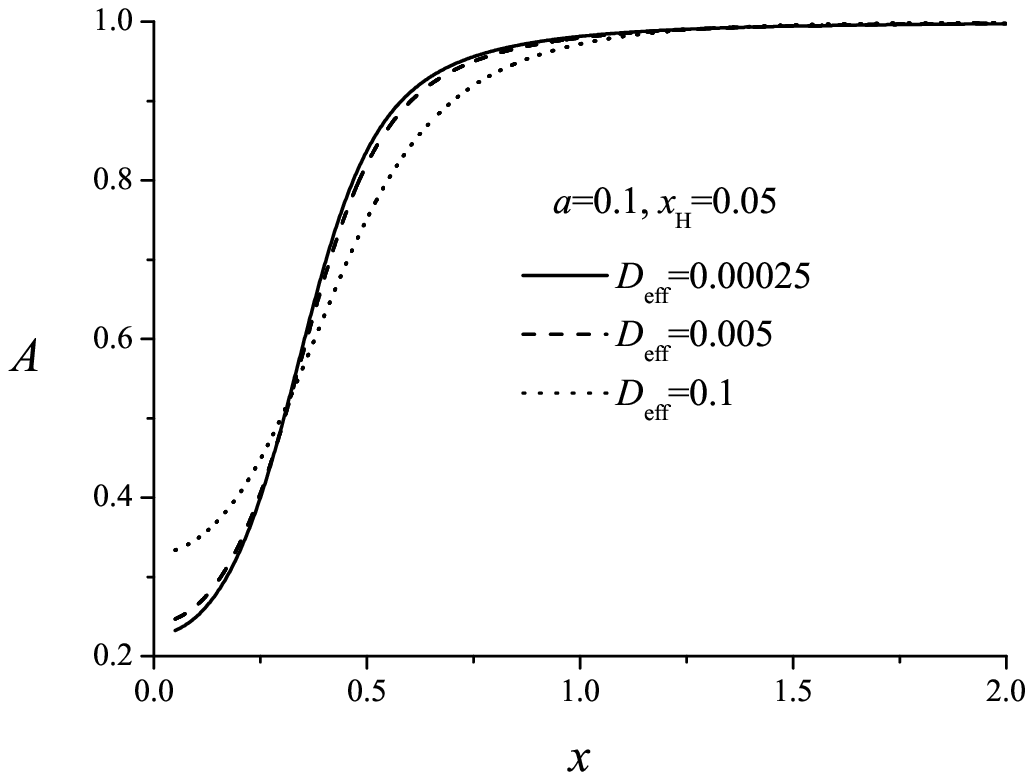}
    \caption{
    The metric functions $\mu(x)$ (left panel) and $A(x)$ (right panel) for black-hole solutions with different values of $D_{\rm eff}$
     ($a=0.15$ and $\tilde{N}=1$). Solutions belonging only to the \textit{lower branch} of solutions are presented
    since the \textit{upper branch} of solutions have similar qualitative behavior.
        }
    \label{fig:mA_BH(x)}
\end{center}
\end{figure}

\begin{figure}
    \begin{center}
    \includegraphics[width=7cm]{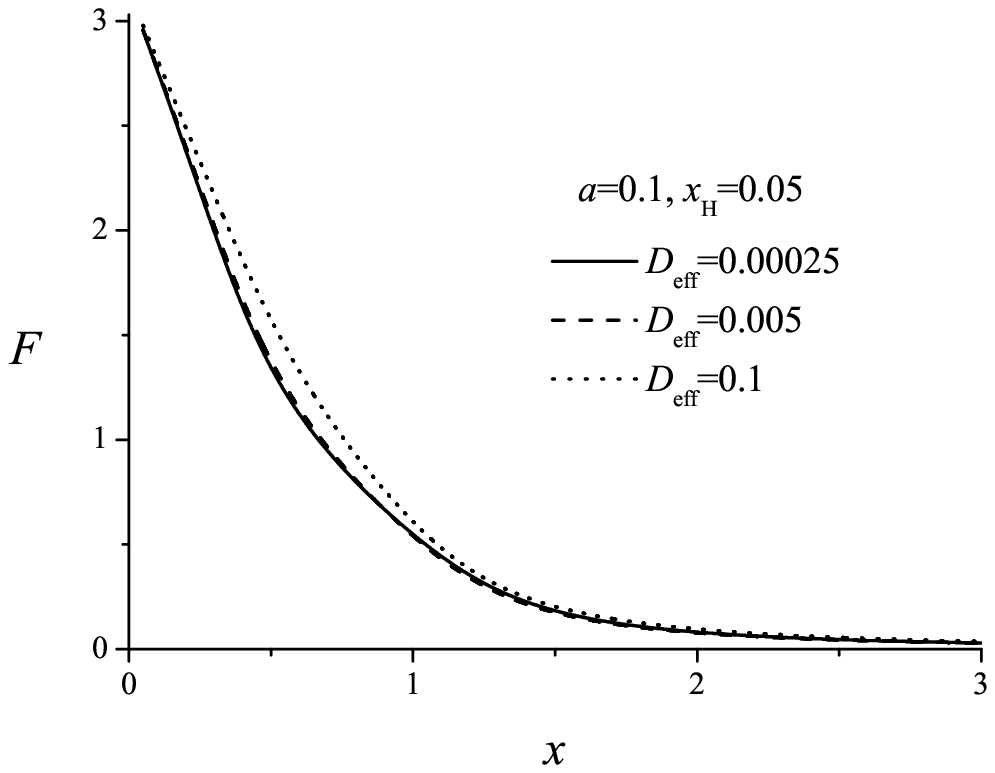}
    \includegraphics[width=7cm]{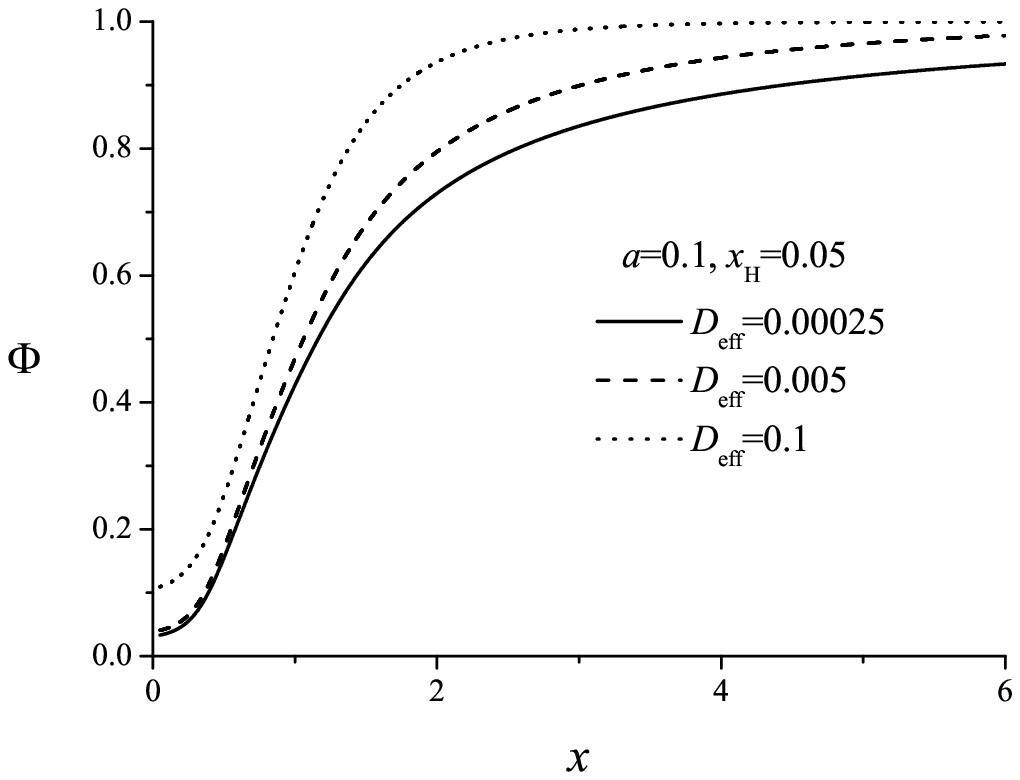}
    \caption{
    The skyrmion field $F(x)$ and the dilaton $\Phi(x)$ for  the same black holes as on Figure \ref{fig:mA_BH(x)}.
        }
    \label{fig:FPhi_BH(x)}
\end{center}
\end{figure}

Like in the ESM \cite{Bizon92}, in the GSM  two branches of black-hole solutions exist which we will denote by
\textit{upper} and \textit{lower branch}\footnote{The notations \textit{upper} and \textit{lower branch} are chosen
using the $F_H(x_H)$ dependences. This is obviously different to the soliton case and actually the \textit{upper
branch} for black holes will have similar properties (such as stability, finiteness/divergency of some of the functions
as $a\rightarrow 0$ ) to the \textit{lower branch} for solitons and vice versa.}. These branches are presented on Figs.
\ref{fig:Frh(a)} and \ref{fig:Phirh(a)} where the shooting parameters $F_H$ and $\Phi_H$ are plotted as functions of
the radius of the horizon  for sequences of black-hole solution when we vary the coupling parameters $D_{\rm eff}$ and
$a$. The metric functions $\mu(x)$ and $A(x)$, the Skrymion field $F(x)$, and the scalar field $\Phi(x)$ are presented
on Figs. \ref{fig:mA_BH(x)} and \ref{fig:FPhi_BH(x)} for several black-hole solutions of the \textit{lower branch} with
$a=0.15$, $r_H=0.05$ and for different $D_{\rm eff}$. The qualitative behavior of these functions is the same for the
\textit{upper branch}.

As it can be seen for fixed values of the coupling parameters,
black-hole solutions exist up to a certain value of the radius of the
horizon $x_H^{max}$.
For fixed values of $D_{\rm eff}$, the maximal radius of the horizon $x_H^{max}$ decreases with the increase of $a$ and eventually reaches
zero at some $a_{\rm max}$ which means that black-hole solutions do not exist for $a>a_{\rm max}$.
The value of $a_{\rm max}$ for which $x_H^{max}$ goes to zero decreases when we increase $D_{\rm
eff}$.

Like in the soliton case, the main difference between the black-hole solutions with and without scalar field is
not qualitative but quantitative.  The values of  $x_H^{max}$ and $a_{\rm max}$ depend
strongly on the parameter $D_{\rm eff}$ and they can be several times larger
in the GSM than in the ESM. Again if we consider $D_{\rm eff}$ close to the one used in
\cite{Nikolaev92},\cite{Nikolaev00} the values of $x_H^{max}$ and $a_{\rm max}$ approach the corresponding values in the ESM.

\begin{figure}
    \begin{center}
    \includegraphics[width=7cm]{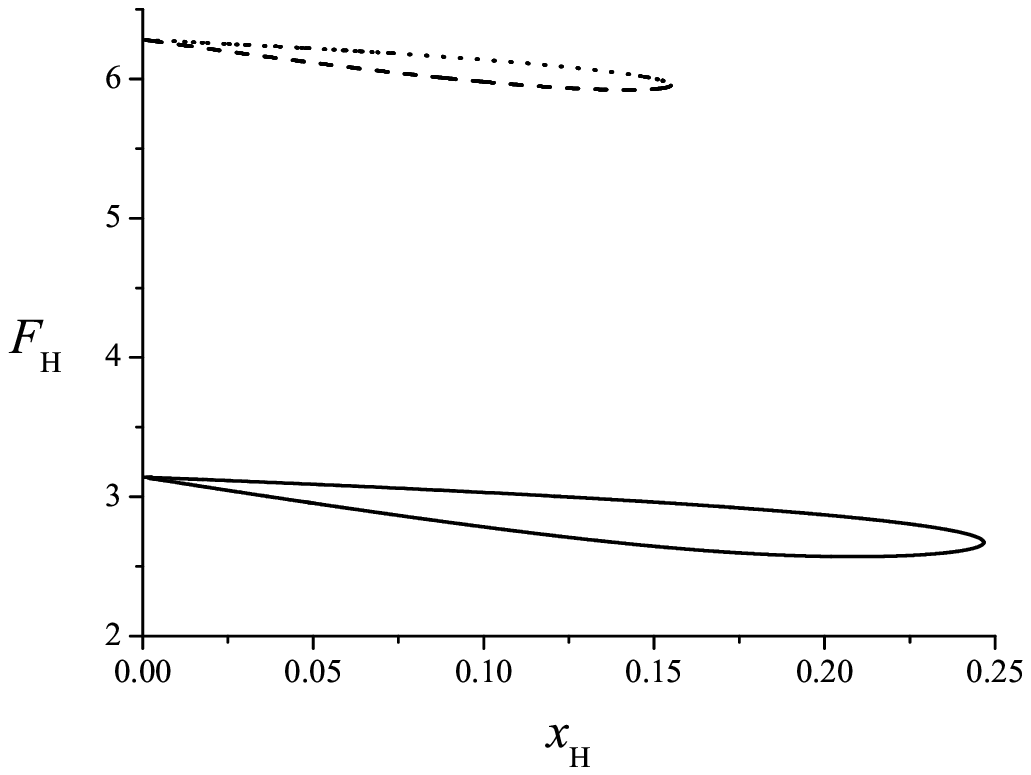}
    \includegraphics[width=7cm]{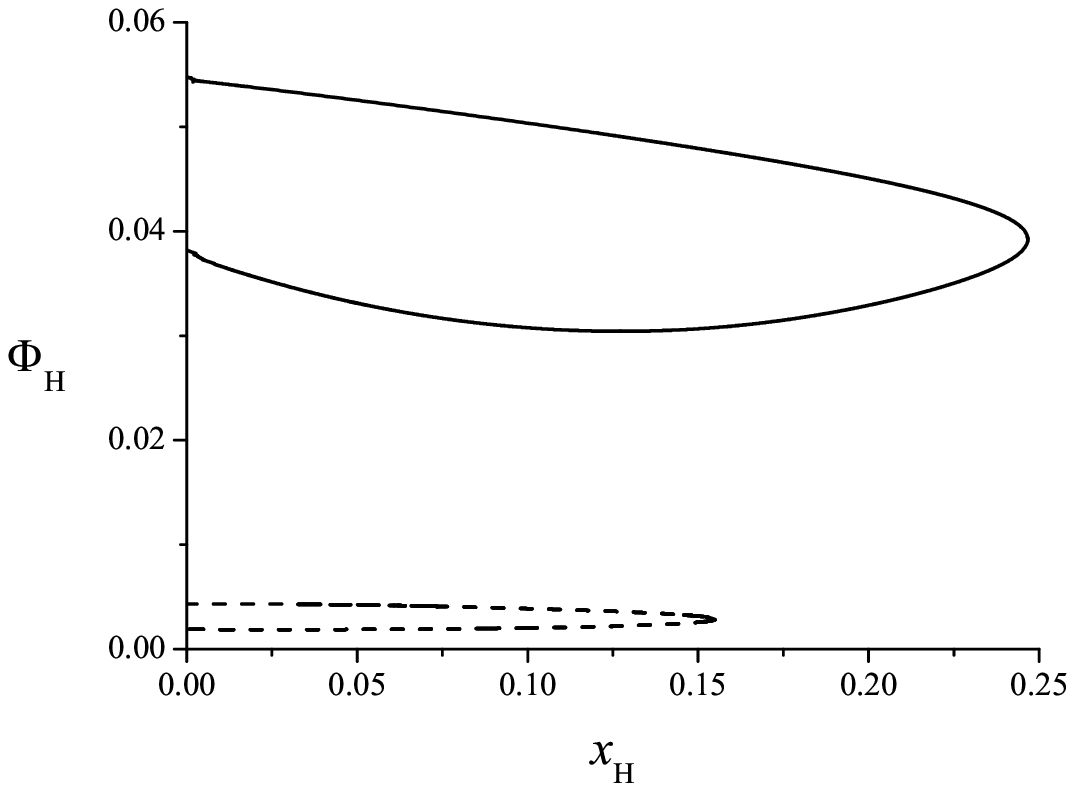}
    \caption{
    The shooting parameters $F_H$ and $\Phi_H$ as functions of the radius of the black-hole horizon $x_H$
    for $D_{\rm eff}=0.00025$ and $a=0.1$. Two sequences of black-hole solutions are shown where the thick line
    corresponds to the same black-hole solutions as the thick line
    on the right panels of Figs. \ref{fig:Frh(a)} and \ref{fig:Phirh(a)}, and the dashed line shows the excited black-hole solutions corresponding to the $n=2$
    soliton solutions.
        }
    \label{fig:FHPhiH_a_var_n}
\end{center}
\end{figure}

There are some interesting properties of the solutions in GSM which are also present in ESM. For fixed $D_{\rm eff}$,
the $a_{\rm max}$ for black holes is the same as $a_{\rm max}$ for the $n=1$ solitons within the numerical error. When
the radius of the horizon goes to zero $x_H\rightarrow0$ the masses of the black holes $M$ and the values of the
shooting parameters $F_H$ and $\Phi_H$ approach the values of $M$, $F_0$ and $\Phi_0$ of the corresponding $n=1$
solitons. The black-hole solutions presented so far correspond to the ground state (baryon number one) solitons. An
infinite series of excitations of both the \textit{upper} and the \textit{lower} branches of solutions corresponding to
the solitons with $n>1$ is also observed. As an example the first excitation is shown on Fig. \ref{fig:FHPhiH_a_var_n}
(the dashed line). As it is expected, the values of $M$, $F_H$ and $\Phi_H$  approach the corresponding values of the
$n=2$ solitons when $x_H\rightarrow0$. The results suggest that an infinite number of excited black-hole solutions
exists and they are in one to one correspondence with the excited soliton solutions with $n>1$.

\section{Stability and thermodynamics of the black holes}\label{Stability_and_thermodynamics}

A very convenient method to infer information for the presence of instabilities in the black-hole solutions is the
Poincare's turning point method \cite{Poincare}. We will apply it here to prove that the \textit{lower branch} of
black-hole solutions is unstable. The method is based on the thermodynamics of the black holes. According to that
method a change of stability is indicated by a turning point or a bifurcation point on the proper conjugate diagram. In
the vicinity of a turning point the branch with a negative slope is unstable. Further details both on the formal proof
and on the application of the method for the study of the stability of compact objects in gravity can be found in
\cite{Sorkin1}--\cite{SYT3}. Some aspects of the thermodynamics of the Einstein-Skyrme black holes have been considered
in \cite{TTM1}.

\begin{figure}
    \begin{center}
    \includegraphics[width=10cm]{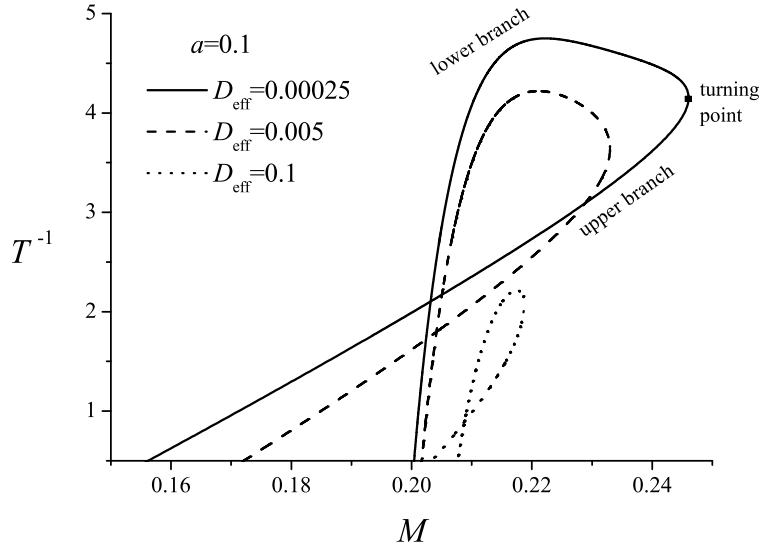}
    \caption{
    The inverse temperature $T^{-1}$ as a function of the black-hole mass $M$ for sequences of black-hole solutions for different
    $D_{\rm eff}$ is shown.}
    \label{fig:Tinv(M)}
\end{center}
\end{figure}

The first law of thermodynamics (FLTD) for the Einstein-Skyrme black holes has been derived in \cite{Zaslavskii92}. A
more  general derivation has been given later in \cite{Heusler93}. For the black holes studied here the FLTD has the
form
\begin{equation}
dM=T \,dS.\label{FL_formula}
\end{equation}
A detailed derivation of (\ref{FL_formula}) is given in Appendix \ref{FL_derivation}. In a micro-canonical ensemble the
equilibrium solutions are the extrema of the entropy $dS=T^{-1}\,dM,$ so the conjugate variables are the mass $M$ and
the inverse temperature  $T^{-1}$. The conjugate diagram $T^{-1}(M)$ is given in Fig. \ref{fig:Tinv(M)} for the
black-hole solutions presented on Figs. \ref{fig:Frh(a)} and \ref{fig:Phirh(a)} (right panels).
It can be seen that the \textit{lower branch} and the \textit{upper branch} merge at a turning point. The slope of the \textit{lower branch} on the
conjugate diagram is negative which, according to the turning point method, means that the branch is unstable. It is
expected that the excited black-hole solutions are dynamically unstable.

The linear stability of both the soliton and the black-hole solutions presented here is currently being investigated.

\section{Conclusion}\label{Conclusion}
In the current paper we have studied new self-gravitating soliton and black-hole solutions in a GSM. The numerical investigation shows that the inclusion of the dilaton does not change the qualitative picture -- it remains the same as in the original ESM. There are some quantitative differences, though.

For both soliton and black holes nonuniqueness of solutions is observed. If we vary the parameter $a$ and keep the rest
of the parameters fixed we obtain two branches of solutions  -- the \textit{upper} and \textit{lower branch}. They
merge at some maximum value $a_{\rm max}$ and no solutions are observed beyond it.  The value of $a_{\rm max}$  of the
solitons coincides with that of the black holes. The same coincidence is observed also in ESM. In GSM, however, $a_{\rm
max}$ is highly dependent on the additional parameter $D_{\rm eff}$ which is related to the dilaton. It increases with
the decrease of $D_{\rm eff}$ and may be several times higher than in the ESM.

For the black holes, if we fix $a$ and vary the radius of the event horizon $x_H$ the two branches merge at some maximal
radius $x_H^{max}$. Beyond $x_H^{max}$ no black-hole solutions exist. Again the quantitative difference between ESM and GSM is that due to the presence of the dilaton $x_H^{max}$ may be considerably higher in the GSM when the same values of $a$ are considered.

In both ESM and GSM with the decrease of the radius of the horizon $r_H\rightarrow0$
the masses of the black holes approach the masses of the solitons from the corresponding branch of solutions.

With the increase of the baryon number an infinite series of solitons and corresponding to them black holes exist
which, however, are expected to be dynamically instable.

In the present paper we also generalize the FLTD of black-hole thermodynamics for the GSM. It has the same form as in
the ESM. From the  application of the turning method we inferred information for the instability of the \textit{lower
branch} of black holes which is in agreement with the situation in ESM.

\vskip 1cm
\appendix

\section{Derivation of the First Law of Thermodynamics}\label{FL_derivation}
For the derivation of the first law of thermodynamics we follow the scheme in \cite{Zaslavskii92}.

Following standard definitions for the Hawking temperature we obtain
\begin{equation}
T=\left.{1\over 4\pi } f\,' A \right|_{r=r_H}= {1\over 4\pi r_H}\, A(r_H)\left(1-{1\over 2}\,v_H+{1\over 2}\,r_H^{\,2}\, V_H\right),
\end{equation}
which together with the entropy of the horizon $S=\pi r_H^2$ gives
\begin{equation}
T {\partial S \over \partial r_H}={1\over 2}\, A(r_H)\left(1-{1\over 2}\,v_H+{1\over 2}\,r_H^{\,2}\, V_H\right).
\end{equation}
The mass of the black holes is given by the formula
\begin{equation}
M=\lim_{r \to \infty} {1\over 2}\, r\,(1-f).
\end{equation}
In order to calculate the variation of the mass we will use the expression \footnote{When $r\rightarrow\infty$,  $r\,f\,A=M+O(r^{-2})$.}
\begin{equation}
r\,f\,A=\int_{r_H}^{r} d r' A(r')\left(1-{1\over 2}\,v+{1\over 2}\,r'^{\,2}\, V\right),
\end{equation}
which can be obtained with the help of eqs. (\ref{eq_m}) and (\ref{eqA}). Then the variation of $M$ is
\begin{equation}
{\partial M \over \partial r_H}=T {\partial S \over \partial r_H}-{1\over 2}\int_{r_H}^{\infty} d r'{\partial  \over \partial r_H}\left[
A(r')\left(1-{1\over 2}\,v+{1\over 2}\,r'^{\,2}\, V\right)\right].
\end{equation}
If the integral in this expression vanishes then
$$
{\partial M \over \partial r_H}=T {\partial S \over \partial r_H}
$$
and the FLTD has the usual form
\begin{equation}
dM=T \,dS. \label{FL}
\end{equation}
The integrable expression can be brought to the form of full derivative with respect to the integration variable which allows the integral to be
estimated. For the derivative in the integral the Leibnitz rule gives
\begin{eqnarray}
&&{\partial  \over \partial r_H}\left[ A\,\left(1-{1\over 2}\,v+{1\over 2}\,r^{\,2}\, V\right)\right]=\notag\\
&&\hspace{1cm}{\partial A  \over \partial r_H}\left(1-{1\over 2}\,v+{1\over 2}\,r^{\,2}\, V\right)+A\,\left(-{1\over 2}\,\,{\partial v \over \partial
\Phi}+{1\over 2}\,r^{\,2}\, \,{\partial V \over \partial \Phi}\right){\partial \Phi  \over \partial r_H}\notag\\&&\hspace{5.6cm}+A\,\left(-{1\over
2}\,\,{\partial v \over \partial F}\right){\partial F  \over \partial r_H}.
\end{eqnarray}
We will treat each of the terms on the right-hand side of the expression above separately. From eq. (\ref{eqA}) we
obtain
\begin{multline}
{\partial A  \over \partial r_H}=A\left\{{u\over r}F\,'{ \partial F  \over \partial r_H}+a\,\tilde{N}\, r\, \Phi\,'{
\partial \Phi  \over \partial r_H}\, +\right. \notag\\+\left.\int_{\infty}^{r} d r'{ \partial F  \over \partial
r_H}\left[{1\over 2 r'}\,F\,'^{\,2}\,{\partial u\over \partial F}-\left({u\,F\,'\over r} \right)'\right] + \right.
\notag\\+\left.\int_{\infty}^{r} d r'{ \partial \Phi  \over \partial r_H}\left[{1\over 2 r'}\,F\,'^{\,2}\,{\partial
u\over
\partial \Phi}-a\,\tilde{N}\left(r'\Phi\,' \right)'\right]\right\}.
\end{multline}

For the expressions in the brackets of the second term and the third term on the right-hand side  from the field
equations for $F$ and $\Phi$ -- (\ref{eqF}) and (\ref{eqPhi}), respectively, we have that
$$
-{1\over 2}\,A\,{\partial v \over \partial F}=- (f\,A\,u\,F\,' )\,'+{1\over 2}\,f\,A\,F\,'^{\,2}\,{\partial u\over \partial F}
$$
and
$$
\,A\,\left(-{1\over 2}{\partial v \over \partial \Phi}+{1\over 2}\,r^{\,2}\,{\partial V\over \partial \Phi}\right)=-a\,\tilde{N}\,(r^2\,f\,A\,\Phi\,'
)\,'+{1\over 2}\,f\,A\,F\,'^{\,2}\,{\partial u\over \partial \Phi}.
$$
Collecting all terms for the integrable we obtain
\begin{equation}
{\partial  \over \partial r_H}\left[ A\,\left(1-{1\over 2}\,v+{1\over 2}\,r^{\,2}\, V\right)\right]=\left(\xi_1\xi_2\right)', \label{integrable_div}
\end{equation}
where
\begin{equation}
\xi_1=r\,f\,A=\int_{r_H}^{r} d r' A(r')\left(1-{1\over 2}\,v+{1\over 2}\,r'^{\,2}\, V\right)
\end{equation}
and
\begin{multline}
\xi_2=\int_{\infty}^{r} d r'{ \partial F  \over \partial r_H}\left[{1\over 2 r'}\,F\,'^{\,2}\,{\partial u\over \partial
F}-\left({u\,F\,'\over r} \right)'\right] +  \notag\\+\int_{\infty}^{r} d r'{ \partial \Phi  \over \partial
r_H}\left[{1\over 2 r'}\,F\,'^{\,2}\,{\partial u\over \partial \Phi}-a\,\tilde{N}\left(r'\Phi\,' \right)'\right].
\end{multline}

From (\ref{integrable_div}) we obtain for the integral
\begin{equation}
\int_{r_H}^{\infty} d r'{\partial  \over \partial r_H}\left[
A(r')\left(1-{1\over 2}\,v+{1\over 2}\,r'^{\,2}\, V\right)\right]=\int_{r_H}^{\infty} d r'\left(F_1F_2\right)'=\left.F_1F_2\right|_{r_H}^{\infty}.
\end{equation}
$\xi_2$ is regular on the event horizon so $\left.\xi_1\xi_2\right|_{r_H}=0$. The situation at infinity is more subtle
since $\left.\xi_1\right|_{\infty}\rightarrow\infty$ and $\left.\xi_2\right|_{\infty}\rightarrow0$. The value of the
integral at infinity can be estimated with the help of the following asymptotic expansion of the functions:
\begin{eqnarray}
&&\left.m(r)\right|_{r \rightarrow \infty}=M-\frac{1}{2}\frac{a F_2^{~2}}{r^3} + {\rm O}\left(\frac{1}{r^4}\right),
~ \left.A(r)\right|_{r \rightarrow \infty}=1-\frac{1}{2} \frac{a F_2^{~2}}{r^4} + {\rm O}\left(\frac{1}{r^5}\right), \label{eq:Assympt_MA_inf}\\
\notag \\
&&\left.F(r)\right|_{r \rightarrow \infty}=\frac{F_2}{r^2}  + {\rm O}\left(\frac{1}{r^3}\right),
~\left.\Phi(r)\right|_{r \rightarrow \infty}=1+\frac{3 F_2^{~2} a}{\tilde{N}\gamma (\varepsilon-4)r^6}  + {\rm O}\left(\frac{1}{r^7}\right),
\label{eq:Assympt_FPhi_inf}
\end{eqnarray}
where $F_2$ is a constant.
With (\ref{eq:Assympt_MA_inf}) and (\ref{eq:Assympt_FPhi_inf}) it can be found that $\left.\xi_1\xi_2\right|_{\infty}\rightarrow0$ which completes
the proof of the FLTD (\ref{FL}).

\section*{Acknowledgements}
This work was partially supported by the Bulgarian National Science Fund under Grants DO No. 02-257, No. VUF-201/06,
and by Sofia University Research Fund under Grant No 88/2011. D.D. would like to thank the DAAD for their support, the
Institute for Astronomy and Astrophysics T\"{u}bingen for its kind hospitality, and Prof. K. Kokkotas, in particular,
for the helpful discussions. D.D. is also supported by the Transregio 7 ``Gravitational Wave Astronomy'' financed by
the Deutsche Forschungsgemeinschaft DFG (German Research Foundation).


\end{document}